# Current eHealth Challenges and recent trends in eHealth applications

Muhammad Mudassar Qureshi[1],MS; Amjad Farooq[1], PhD; Muhammad Mazhar Qureshi[2], MPH

[1]University of Engineering and Technology, Lahore, Pakistan
[2]World Health Organization

Corresponding Author:
Muhammad Mudassar Qureshi, MS
University of Engineering and Technology
G.T. Road
Lahore, Punjab
Pakistan
Phone: 92 300 4889 526
Email: mudassar66@gmail.com

**Abstract**:

eHealth (Health Informatics/Medical Informatics) field is growing worldwide due to acknowledge of reputable Organizations such as World Health Organization, Institute of Medicine in USA and several others. This field is facing number of challenges and there is need to classify these challenges mentioned by different researchers of this area. The purpose of this study is to classify different eHealth challenges in broader categories. We also analyzed recent eHealth Applications to identify current trends of such applications. In this paper, we identify stakeholders who are responsible to contribute in a particular eHealth challenge. Through eHealth application analysis, we categories these applications based on different factors. We identify different socio-economic benefits, which these applications can provide. We also present ecosystem of an eHealth application. We gave recommendations for eHealth challenges relevant to Information Technology domain. We conclude our discussion by specifying areas for future research and recommending researchers to work on identify which type of disease can control and manage by different eHealth applications.

# 1 Introduction

eHealth refers to use of information and communication technologies to improve or enable health and health care (Nykänen, 2006). eHealth is also referred as medical informatics and Health Informatics (Pagliari, 2007). Many eHealth systems are developed (Iliakovidis, 2004.) in last decade. In last few years, field of eHealth has grown worldwide (WHO, National eHealth strategy toolkit, 2012). The Institute of Medicine (U.S.A) has acknowledged that Information technology can play important tool to improve healthcare cost, patient safety and quality in medical care (Kohn LT, 2000;America, 2001).In (Budych, 2014), authors suggest that eHealth can play important role to meet future challenges in field of Health Care. Application of information technology in health starts in mid 90s (Doupi, 2005). Main target areas of eHealth are 1) managing electronic Health Records; 2)Communication infrastructures and networks; 3) Standardization of patient's data; 4)Security and Privacy;5)Research, national and international collaboration. Applications, which are related to Citizen and Patients, are core elements of eHealth. Such applications are in strategies of some countries like Denmark, Germany, Iceland, Ireland and U.K (T, 2000).Denmark and Sweden are leading this field. In Denmark, 70% prescription transmitted electronically and In Sweden, transmitted prescription percentage is 45.There are several advantages of developing eHealth systems. Some key advantages of eHealth systems are:

- Health Care Systems- optimize resource use (quality – cost effectiveness)
- Facilitate better decision-making.
- Information and data: Increased accessibility, availability, Speed
- Impact on Professionals: optimize performance (accuracy, knowledge expertise)
- Helpful in evaluating essential data of patient's identity and medical history.

There are number of eHealth systems developed but many promises of eHealth research and development not fulfill yet (Doupi, 2005). Effective and efficient development of eHealth system faces number of challenges. Most Important issues are lack of commitment by healthcare authorities and missing interoperability among different health information systems. From past more than a decade, researchers are raising different challenges facing by eHealth. There is need to review these challenges and classify in different categories.

The purpose of this paper is to classify eHealth challenges in broader categories by extensive review of articles published in this area. We classify articles based on different parameters. We also identify stakeholders who may responsible to contribute in a specific eHealth challenge. This reveals that, almost half of challenges are relevant to Information Technology (IT) experts. Based on this fact, we propose solutions to eHealth challenges related to IT domain. We also analyzed recent eHealth applications to identify current trends in this area. We categorized eHealth application based on different factors. Purpose of this categorization is to identify how much work done to control and manage a particular disease. And which type of disease is more suitable for developing an

eHealth application. Next, we will discuss current challenges in eHealth. Then, we presented recent eHealth applications. After that, we present methodology how we review different research articles and how we analyzed eHealth applications. Next, we discuss results. In section 5,we give recommendations for different eHealth challenges relevant to IT and recommendations related to eHealth Applications. In last section, we conclude our discussion and recommend future research areas.

# 2 Current Challenges in eHealth

## 2.1 Detection of disease at early stage

Detection of disease at early stage helps not only to reduce cost of medical treatment but it is also useful in saving valuable lives of people. For instance, detection of cancer at early stage may rescue man's life instead of detection of disease at later stage. One significant challenge is that there is no infrastructure available for detection of disease at its early stage (TanJent, 2006;Paul H Keckley, 2010). Interactive Health Communication Applications (IHCAs) can be used for overcoming this challenge(Casey, 2014). In(Iluz, 2014), author discussed algorithm, which can be use in early detection of Parkinson's disease.

## 2.2 Reducing Cost of Health Care by using eHealth System

Reducing Cost of Health care with help of eHealth is a big Challenge (Atienza, 2007;Kostkova, 2015;Fosso Wamba, 2013TanJent, 2006;). HIEs (Health Information Exchanges) is one module in developed Health Care Systems. Purpose of this module is to get access of complete view of treatment plans of their patients.

## 2.3 Efficient Managing Patient's Data in eHealth System

Capturing, Storing and maintaining data and Accessing Information in efficient way is also a big challenge (TanJent, 2006). Efficiently maintaining EHR (Electronic Health record) is a big issue (Atienza, 2007). There is need of clear data standards to get optimal value in implementing eHealth Systems (WHO, National eHealth strategy toolkit, 2012).

## 2.4 Effective utilization of Skills of HSR and IT Expert

Health Service Researcher (HSR) is a person who is focal person in development of eHealth Solution. His time is precious and saving his time by minimal involvement in development of eHealth solution is a challenge. So, combining skills of both HSR and IT Expert in an effective manner to achieve maximum benefit is a great challenge (Pagliari, 2007; Association., 2012).

## 2.5 Establish trust between HSR and IT Expert

As discussed above, HSR is main person who is involved in development of eHealth solution along with IT experts. So, trust should be established between both HSR and IT Expert team is necessary but issues arise when both HSR and IT expert are interacting with each other. Therefore, establishing a trust and mutual respect between both HSR and IT expert team is a big challenge. (Pagliari, 2007; Association., 2012)

## 2.6 Patient Privacy

Privacy of patients is challenge in development of eHealth Systems (Kostkova, 2015). In (Dong, 2012), authors suggested two key privacy challenges: enforcing privacy and privacy in the presence of others. In (ALKRAIJI, 2014), authors also convinced that privacy is a key issue in GCC (Gulf Cooperation Council) Countries.

To overcome this issue, legal regulation is needed otherwise resistance will face to adopt eHealth system(Ajami S, 2013). We have to ensure that Patient have trust on system that their personal information is protected (AV, 2007;Chalmers D, 2004;Sass M, 2011).

## 2.7 Interoperability of data among different Health Care Places

Patient's data inter operability among different Health care Places like Hospitals, private clinics is a key issue (Nagai, 2012; SJ. 2007; Viswanath K, 2007;ALKRAIJI, 2014). Due to lack of interoperability, data remain fragmented, isolated and data analysis cannot be done (Glaser, 2011). Due to this issue, exchanging data among different systems is not possible which is hindrance to accomplish fundamental goals of healthcare(Sass M, 2011).WHO also recommended its members to adopt standards for effective exchange of information between eHealth implementations and health care practitioners (WHO, 2014). Solution to this problem is making data in a standard form (Hammond WE, 2010).

## 2.8 Shortage of Professionals

Researchers and professionals are fewer in this interdisciplinary area in GCC Countries (ALKRAIJI, 2014). However, there is shortage of such professionals in all part of the world (DE., 2010;Hersh W, 2010;Qureshi, 2014).

In order to overcome this issue, e-Capicity meeting in Bellagio was held in 2008 (Foundation., 2010). It was decided to give training (both formally and un formally) to health workers which increase their skills, attitude and level of knowledge related to health informatics.

Also, there is need of specialized degree programs in this discipline. AMIA's (American Medical Informatics Association's) took initiative to train 10,000 professionals in Health Informatics (HI) by 2010 (Hersh W W. J., 2007). This association worked in developing countries like Singapore and Argentina to create an international version adapted to local needs (Otero P, 2007;Margolis A, 2013).

Another approach to address this problem is use of mobile and telemedicine devices to connect trained resources with population. It is especially useful in rural areas. Such initiative is taken in India where mobile tools are used for screening of retinopathy (Murthy KR, 2012).

## 2.9 Complexity of Health Care Infrastructure

It is a big challenge to manage complexity of Healthcare infrastructure. (ALKRAIJI, 2014;Nagai,2012) Healthcare infrastructure can be complex due to different reasons. Populated areas (China, India and other populated areas of developing countries) have to develop many hospitals, many Health care centers. Similarly geographically dispersed areas have also very complex Health Care infrastructure.

Such complex infrastructure should support eHealth but currently support to eHealth is insufficient and not distributed well (Luna, 2014). There are many factors for such difficulties. For example in sufficient support of electricity (Latourette MT, 2011), poor quality or not availability of Internet access (Shiferaw F, 2012). These problems are more common in rural areas (Simba DO, 2004;Fraser HS, 2004).

However, mobile phone infrastructure is developing at an increasing rate (Lewis T, 2012) provides opportunities to implement systems with less resources (E., 2009). Due to this factor mobile health (part of broader telemedicine field) can be useful in presence of these inadequate infrastructure (Asangansi I, 2010). However these solutions can lead to other problems such as fragmented information and difficulties for project scalability.

Hardware and Software are integral part of eHealth System infrastructure. It is our fortune that now Hardware cost is drastically reduced as compare to forty years before (GE., 2006). Due to low cost of Hardware developing and under developing countries are in position to make initiative of distributing low cost Computers (ceibal, 2014;Igualdad,2014;Rwanda,2014). Open Source movement is helping limited resource countries in terms of Software. PostgreSQL (a open Source DBMS) and OpenMRS (Helps to design customized EHRs) are two good examples of Open Source Software (PostgreSQ,2014;Mamlin BW, 2006). OpenMRS is implemented in many developing countries of Africa, Asia and Central and Latin America (Mohammed-Rajput NA, 2011;Gerber T, 2010). In most cases, such programs are implemented in resource-limited countries with help of donor funding (Gordon AN, 2007) for initial stages. For running and scaling up such programs need on going finance which is difficult for such countries (Lewis T, 2012).

### 2.10 eHealth friendly Government Policies

eHealth friendly policies are not developed in most part of the world. In Japan, eHealth friendly policies are not being developed because policy makers have less exposure to eHealth discipline and potential benefits of this area(Nagai, 2012). Similarly in GCC countries, policies are not friendly for eHealth (ALKRAIJI, 2014). Also in Jordan, government policies are not suitable for development of eHealth (Matar, 2014).

There is need of development of eHealth System framework for better development as well as sustainability of eHealth Projects. It is advocated by my reputable international organizations such as United Nations (UN), World Health Organization (WHO) (Bank., 2006).

### 2.11 Other Issues

In (Matar, 2014) authors arguments that people are resistant towards advancement in technology, which involve in eHealth. Top issues of Ghana are Lack of ICT Infrastructure, Basic ICT Knowledge/Skills, Internet accessibility, Financial & sustainability issues (Bailey, 2015). Involving patients and their relatives while caring a patient when he is suffering from disease or in process of rehabilitation is also an issue (Bedeley, 2014;Gard, 2012).

## 3 Current Health Applications

We analyzed several Health related Applications and based on that analysis, we categorize Health Applications in three broader Categories

- Health Information Systems
- Serious Games
- eHealth Applications

Health Information Systems cannot call as eHealth Application. For details see section 3.1. Serious Games and eHealth Applications discussed in coming sections can called as eHealth application as these two type of applications are developed to solve any Health issue(s).

Case Management is a process in Health in which any patient comes to a Physician for his/her treatment. Physician does many activities during patient's treatments. Major activities are monitoring, detection, treatment and rehabilitation. So, all these activities are part of Case Management. Any issue in Case Management can also called Health Issue. Lack of training of physicians or unawareness of General Health education (lack of knowledge to prevent from a disease) are also issues in Health, which can be address by eHealth Applications.

We can Categorize Health Issues based on several Health Management Processes. These Processes are listed as follows:

- Monitoring
- Detection/Diagnosis

- Treatment
- After Treatment / Rehabilitation
- Physicians Training
- General Health Education Awareness

We discussed each Health application category in detail in following sub sections.

### 3.1 Health Information Systems

Health information systems are systems that are developed for different Health related Activities. Hospital Management, Radio Information System, Patient Record Management Systems, Telemedicine or Tele Consultation Systems are some type of Health Information Systems. Such systems are developed not to fulfill any eHealth Challenge discussed in Section 2.

### 3.2 Serious Games

Different types of games are created. Each game has its own objectives. Some games objective is only to amuse its target clients. As a side effect, they can help to solve any health issue. Some games are two folded objectives. One objective is to give entertainment and other is to share knowledge. There are also some games which objective is to help to solve any medical problem. Mostly, such games are simulation games. So, we are dividing games in three categories basis on their objectives and we list these categories as follows:

- Recreational Games
- Hybrid Games (Recreational and Problem Solving)
- Pure Problem Solving Games

Now, we will introduce some important games in each sub section.

#### 3.2.1 Recreational Games

As discussed above, this type of games main objective is to provide entertainment. But as a side effect, some health issues may address. (Game, 2015)Dance Dance Revolution (DDR) Game is an entertainment game but as a bonus, due to body exercise ; it will help to maintain good health and avoid problems due to not doing any physical exercise. An exercise game WII (Rego P, 2010) is also a popular game and can be use to maintain good health.

#### 3.2.2 Hybrid Games

These games serve both purposes (entertainment an Health Solution). FatWorld (Sliney A, 2008), Re-Mission (Vidani AC, 2010), Air Medic Sky (Vtnen A, 2008) are most commonly used games.

#### 3.2.3 Pure Problem Solving Games

These types of games are developed to solve specific health problem. Some simulation games are Virtual Dental Implant Training Simulation Program(BreakWayGames, 2015), EMSAVE (HCI Lab, 2011)and Olive: 3d Hospital Training(Scarle S, 2011).

Several applications are related to Case Management. Like, some Monitoring games are CHF Tele-management System (Finkelstein J, 2010), Health Care Monitoring(Fergus, 2009) and U-Health Monitoring System(Lee, 2009).
Unobtrusive Health (McKanna, 2009)and EEG-Base Serious Games(Wang, 2010) are related to detection process.
Some Treatment Process related games are Match-3(Scarle, 2011), Diagnosis and Management of Parkinson(Atkinson, 2010) and Social Skills(Bartolomé, 2010). Some applications are related to Rehabilitation process. Some such applications are ULRFS(Burke, 2009)and After Parkinsons Disease(Studios, 2015). (Boulos, 2015) Monster Manor is a game app for childen having type 1 diabetes. (Boulos, 2015) Empower is a game app for childen having type 1 diabetes. (Boulos, 2015) Health Seeker is a game app for patients having type 2 diabetes. It is Facebook application.

Some professional training applications are HumanSim(Associates, 2015), Virtual Dental Implant Training Program(BreakWayGames, 2015), Nursing and Midwifery(Skills2Learn, 2015), Pulse(BreakWayGames, 2015), MUVE(J, 2010), Game Based Learning for virtual patient(IFoMaICL, 2008), Medical Simulation Training Program(Sliney, 2008), VI-MED(Mili, 2008) and Coronaryartery bypass surgery procedure Serious Game (Sabri H, 2010).

### 3.3 eHealth Applications

eHealth applications are application that are developed purely to solve any Health related issue or to facilitate any health process. We discussed different eHealth application in this section.

(Pan, 2015)Mobile application for Parkinson's disease Monitoring(PD Mobile App)is developed to monitor and diagnose different behaviors of PD infected patient.

(Semple, 2015)Mobile App for monitoring of Post Operation is a mobile app for monitoring of post operation. (Ho, 2015)A tele-surveillance System (TSS for ECG) is developed to monitor electrocardiodiagram(ECG). (Arden-Close, 2015)A visualization tool is used to help users to lose their weights (VT for Weight Loss). (Huguet, 2015)myWHI is a mobile application, which is used to monitor headache in young Adults.(Lim, 2015) PotM is a mobile application helps to discover a disorder called Pre-eclampsia. (Ferrando, 2015) Sintromacweb is tool to manage OAT(Oral Anticoagulation Therapy).(Zan, 2015) iGetBetter System is used by Patients for self-managing disease of heart failure. In (Henriksson, 2015), author discussed TECH(Tool for energy balance in children) to monitor energy intake.(Volker, 2015) eHealth module embedded in collaborative Occupational Health Care (ECO ) is tool for rehabilitation of patient of common mental disorder.(Cristancho-Lacroix, 2015)Diapason is Web Based physcoeducational program for informal care givers of persons with Alzheimers disease.(Yoong, 2015) A web based program to implement healthy eating and physical training policies (WPHTP) is used to educate public. (Williams, 2015) Dynamic Consent is a tool use to increase trust of patients to share their electronic patient records publically.(Rodriguez, 2015) e-consult tool is developed for health care of Veteran Affairs.(English, 2015)FAIR is a tool which is use in assiting family based research in existing data warehouse.(Singh, 2015) A tool is developed to assist physicians to read radiology reports.(Huckvale, 2015) Different smart phone apps are discussed for calcualting Insuline dose. (Hardinge, 2015) A mobile application used for monitoring and treat ment for COPD (Chronic Obstructive Pulmonary Disease ).

# 4 Methodology

For identifying different challenges in eHealth, we adopt approach similar to these authors.(Ngai, E. W. ,2002;Ngai, E. W. ,2009;Fosso Wamba, 2013; Luna,2014). Some authors only choose Journals articles for their research (Ngai, E. W. ,2002;Ngai, E. W. ,2009;Fosso Wamba, 2013) and totally exclude internatioanl conferences , reports and web references. Some authors uses different research databasesusing key words related to their research area(ALKRAIJI, 2014;Luna,2014).In our research, our first choice is Journal articles and most of articles which we studied are journal articles, however we also consider some articlesof recognized research conferences and a few reports of reputed Organizations.

We classified eHealth challenges in 10 different categories 1)Detection of disease at early stage; 2)Reducing Cost of Health Care by using eHealth System; 3)Efficient Managing Patient's Data in eHealth System; 4)Effective utilization of Skills of HSR and IT Expert; 5)Establish trust between HSR and IT Expert; 6)Patient Privacy;7)Interoperability of data among different Health Care Places;8)Shortage of Professionals;9)Complexity of Health Care Infrastructure;10)eHealth friendly Government Policies.

For eHealth application, we mostly analyze eHealth applications, which are developed in current year or in last year. We categorize these applications in two categories and analyzed based on Health Management Processes.

# 5 Results

This section is divided in two sub sections. In section 5.1, we discussed results by reviewing different research articles related to eHealth Challenges. Section 5.2 showed results of reviewing different eHealth Applications.

## 5.1 Results of reviewing different Research Articles related to eHealth Challenges

First, we identified different stakeholders who are responsible to address different eHealth challenges. Next we showed distribution of articles by year. Then, we categorized articles by different challenges. After that, we presented article classification based on article publication per Journal.

### 5.1.1 Stakeholders responsible for addressing different eHealth challenges

We classified eHealth challenges into different stakeholders who are responsible to do necessary actions to meet these challenges. Because it is an interdisciplinary area, so not one stakeholder can resolve these issues, however as presented in Table 1, IT Experts can do research and can provide solutions to at least 5 eHealth challenges. So purpose to classify these challenges is to show how different stakeholders can participate to their relevant area.

We can also infer that it is wide-open area of research related to IT researchers who can contribute a lot in this area.

| **eHealth Challenge** | **Stakeholder** |
|---|---|
| Detection of disease at early stage | IT Experts |
| Reducing Cost of Health Care by using eHealth System | IT Experts |
| Efficient Managing Patient's Data in eHealth System | IT Experts |
| Effective utilization of Skills of HSR and IT Expert | IT Experts, Government, NGOs |
| Establish trust between HSR and IT Expert | Government, NGOs |
| Patient Privacy | Government, NGOs |
| Interoperability of data among different Health Care Places | IT Experts |
| Shortage of Professionals | Government, NGOs, Educational Institutes |
| Complexity of Health Care Infrastructure | Government, NGOs |
| eHealth friendly Government Policies | Government, NGOs |

**Table 1: eHealth Challenges Classification based on Stakeholders**

### 5.1.2 Articles distribution by publication year

Based on our literature review, we founded there is no publication in 2003,2005 and 2008. There was one publication in 2002 and 2009. In 2007 and 2010, maximum number of articles published which were 8.This trend is shown in Figure 1.

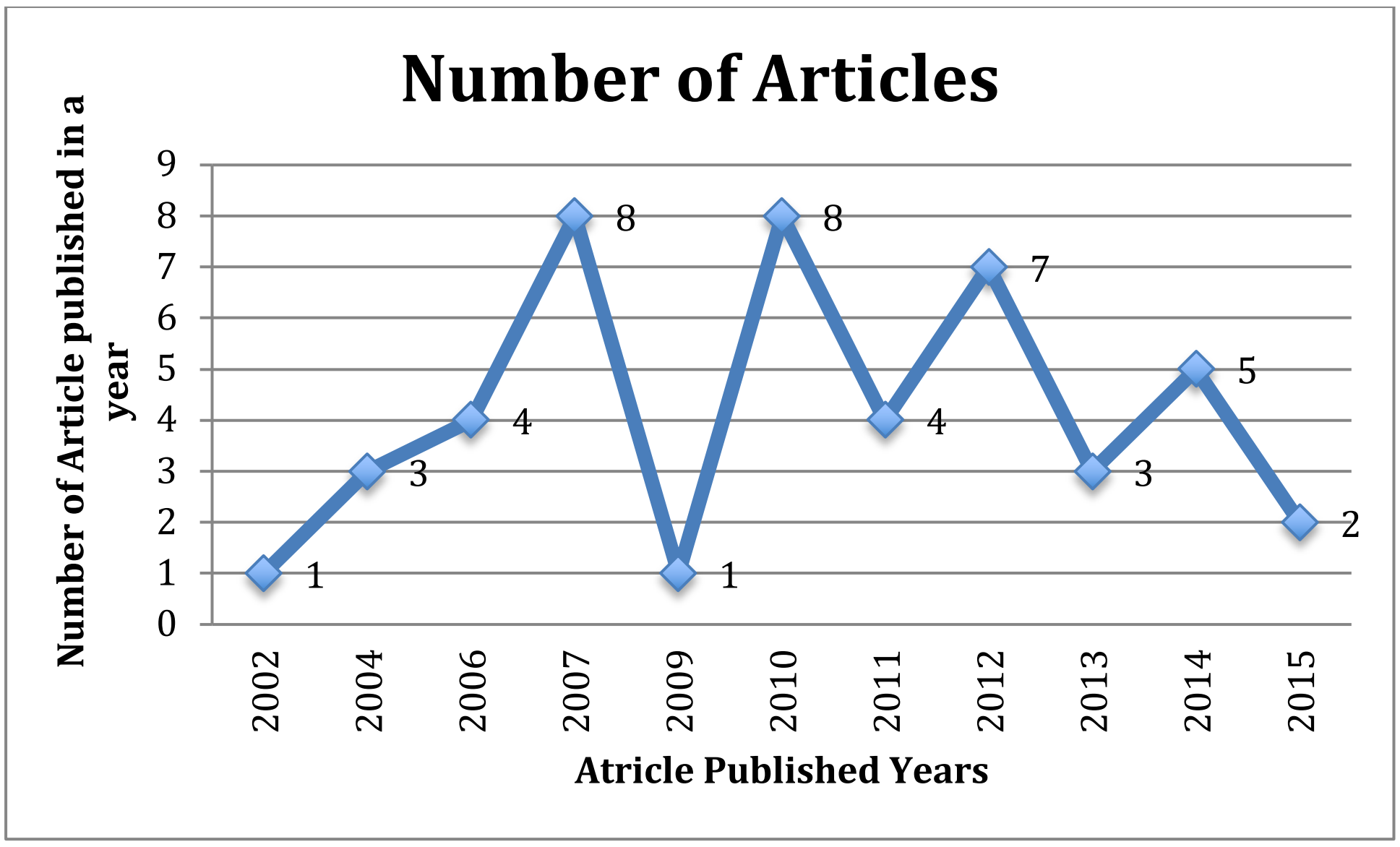

Fig.5.1: Articles Published per Year

### 5.1.3 Articles distribution by eHealth Challenge

Table 2 shows distribution of articles for each eHealth challenge. This may useful for researcher who wants to research on specific eHealth issue to realize the importance of that topic.

| eHealth Challenge | Number of Articles |
|---|---|
| Detection of disease at early stage | 3 |
| Reducing Cost of Health Care by using eHealth System | 3 |
| Efficient Managing Patient's Data in eHealth System | 3 |
| Effective utilization of Skills of HSR and IT Expert | 2 |
| Establish trust between HSR and IT Expert | 3 |
| Patient Privacy | 7 |
| Interoperability of data among different Health Care Places | 7 |
| Shortage of Professionals | 10 |
| Complexity of Health Care Infrastructure | 20 |
| eHealth friendly Government Policies | 3 |

Table2: Number of Articles per Challenge

### 5.1.4 Classification of articles published in a journal

Table 3, classified number of article published in one Journal. American Journal of Preventive Medicine and Health Affairs (Millwood) have maximum number of articles relevant to eHealth challenges.

| Journal Name | Number of Articles |
|---|---|
| African Health Sciences | 1 |
| American Journal of Preventive Medicine | 3 |
| British Medical Journal | 3 |
| Diabetes Research and Clinical Practice | 1 |
| Glob Health Action | 2 |
| Frontiers in public health | 1 |
| Health Affairs(Millwood) | 3 |
| Healthcare informatics research | 2 |
| IEEE | 1 |
| International Arab Journal of e-Technology | 1 |
| International Journal of Biotechnology | 1 |
| International Journal of Health Care Quality Assurance | 1 |
| International Journal of Information Management | 2 |
| International Journal of Medical Informatics | 1 |
| Journal of Digital Imaging | 1 |
| Journal of the Healthcare Financial Management Association | 1 |
| Journal of Health Informatics in Developing Countries | 1 |
| Journal of medical Internet research | 1 |
| Journal of Telemedicine and Telecare | 1 |
| King's Law Journal | 1 |
| Telemedicine journal and e-health : the official journal of the American | 2 |
| The Bulletin of the World Health Organization | 1 |

Table 3: Classification of Articles per Journal

## 5.2 Results related to Current eHealth Applications

Firstly, we classified eHealth related serious games into Health Management Processes and showed that which game is developed for which specific Health Management Process(s). Next, we categorized eHealth Applications based on Health Management Processes. In Section 5.2.3, we presentedeHealth Applications distribution based on Health Management Processes.

### 5.2.1 Classification of eHealth Serious Games by different Health Management Processes

In Table 4, we distributed different eHealth related serious games based on Health Management processes.

| | Monitoring | Diagnosis | Treatment | Rehabilitation | Training for Professional | Training / Heath Awareness for Non Professional |
|---|---|---|---|---|---|---|
| DDR(Game, 2015) | | | | | | √ |
| WII (Rego P, 2010) | | | | | | √ |
| FatWorld (Sliney A, 2008), | | | | | | √ |
| Re-Mission (Vidani AC, 2010) | | | | | | √ |
| Air Medic Sky (Vtnen A, 2008) | | | | | | √ |
| Virtual Dental Training Program (BreakWayGames, 2015) | | | | | √ | |
| EMSAVE (HCI Lab, 2011) | | | | | √ | |
| Olive: 3d Hospital Training (Scarle S, 2011) | | | | | √ | |
| CHF Tele-management System (Finkelstein J, 2010) | √ | | | | √ | |
| Health Care Monitoring (Fergus, 2009) | √ | | | | | |
| U-Health Monitoring System (Lee, 2009) | √ | | | | | |
| Unobtrusive Health (McKanna, 2009) | | √ | | | | |
| EEG-Base Serious Games (Wang, 2010) | | √ | | | | |
| Match-3 (Scarle, 2011) | | | √ | | | |
| Diagnosis and Management of Parkinson (Atkinson, 2010) | | | √ | | | |
| Social Skills (Bartolomé, 2010) | | | √ | | | |
| ULRFS (Burke, 2009) | | | | √ | | |
| After Parkinsons Disease (Studios, 2015) | | | | √ | | |
| (Boulos, 2015) Monster Manor | √ | √ | √ | | | |
| (Boulos, 2015)Empower Manor | √ | √ | √ | | | |
| (Boulos, 2015)Health Seeker | √ | √ | √ | | | |
| HumanSim (Associates, 2015) | | | | | √ | |
| Virtual Dental Implant Training Program (BreakWayGames, 2015) | | | | | √ | |
| Nursing and Midwifery (Skills2Learn, 2015) | | | | | √ | |
| MUVE (J, 2010) | | | | | √ | |
| Game Based Learning for virtual patient (IFoMaICL, 2008) | | | | | √ | |
| Medical Simulation Training Program (Sliney, 2008), | | | | | √ | |
| Medical Simulation Training Program (Sliney, 2008) | | | | | √ | |
| VI-MED (Mili, 2008) | | | | | √ | |
| Coronary artery bypass surgery procedure (Sabri H, 2010). | | | | | √ | |

Table 4: Classification of eHealth Serious Games based on different Health Management processes

### 5.2.2 Classification of eHealth Applications by different Health Management Processes

We classified recent eHealth Applications based on Health Management processes in Table 5.

| | Monitoring | Diagnosis | Treatment | Rehabilitation | Training for Professional | Training / Heath Awareness for Non Professional |
|---|---|---|---|---|---|---|
| PD Mobile App (Pan, 2015) | √ | √ | | | | |
| Monitoring of Post Operation (Semple, 2015) | | | | √ | | |
| TSS for ECG(Ho, 2015) | | | √ | | | |
| VT for Weight Loss (Arden-Close, 2015) | | | | | | √ |
| myWHI (Huguet, 2015) | √ | √ | √ | | | |
| PotM JMIR mHealth and uHealth 3.2 (2015). | | √ | | | | |
| Sintromacweb (Ferrando, 2015) | | | √ | | | |
| Tech(Henriksson, 2015) | √ | | | | | √ |
| ECO (Volker, 2015) | | | | √ | | |
| (Cristancho-Lacroix, 2015) | | | | | | √ |
| WPHTP(Yoong, 2015) | | | | | | √ |
| Dynamic Consent(Williams, 2015) | √ | | | | | |
| (Rodriguez, 2015) | √ | √ | √ | | | |
| FAIR (English, 2015) | | √ | | | | |
| (Singh, 2015) | | √ | | | √ | |
| Huckvale, 2015 | √ | | | | | √ |
| Hardinge, 2015 | √ | | √ | | | √ |

Table 5: Classification of eHealth Applications based on different Health Management processes

### 5.2.3 Distribution of eHealth Applications (Both Serious games and eHealth Application) by different Health Management Processes

In this section, we showed how many eHealth applications (both eHealth serious games and other eHealth Applications) are developed against each Health Management Process.

| Health Management Process | Number of eHealth Applications |
|---|---|
| Monitoring | 13 |
| Diagnosis | 11 |
| Treatment | 10 |
| Rehabilitation | 15 |
| Training for Professionals | 15 |
| Training for Non-Professionals | 11 |

Table 6: Distribution of eHealth Applications by different Health Management processes

### 5.2.4 Classification of Health Applications based on Disease Control and Management

Doing exercise to prevent from disease can called as curing. Curing is a disease control technique. Any simulation program related to management of any disease or overall Health Management System also helpful in improving quality of Health Management. Such systems help inDisease Management. Any eHealth application used to help in managing any disease is also helpful in improving Health Management Quality. Based on these, we classified Health Applications based on Disease control and management in following two tables. In Table 7, we gathered all applications, which are related to training, simulations or body fitness.

| Health Application(s) | Health Application Type | Training/Simulation |
|---|---|---|
| **DDR (Game, 2015), WII (Rego P, 2010), Re-Mission (Vidani AC, 2010)** | Serious Games | Games related to Body Fitness |
| **Air Medic Sky (Vtnen A, 2008)** | Serious Games | Related to Doctors training of new techniques |
| **EMSAVE (HCI Lab, 2011)** | Serious Games | It is virtual training environment for emergency medical care |
| **Olive: 3d Hospital Training (Scarle S, 2011), Health Care Monitoring (Fergus, 2009), U-Health Monitoring System (Lee, 2009), HumanSim (Associates, 2015)** | Serious Games | Training applications related to hospital management |
| **Nursing and Midwifery (Skills2Learn, 2015), VI-MED (Mili, 2008)** | Serious Games | Training applciations for Nursing |
| **Game Based Learning for virtual patient (IFoMaICL, 2008), MUVE (J, 2010)** | Serious Games | Simulation applications using virtual patients |
| **Medical Simulation Training Program (Sliney, 2008),** | Serious Games | Application for medical education learning |
| **Pulse (BreakWayGames, 2015)** | Serious Games | Application for training Young health care professionals clinical skills |
| **Virtual Dental Implant Training Program(BreakWayGames, 2015)** | Serious Games | Dental related Diseases |
| **Unobtrusive Health (McKanna, 2009)** | Serious Games | Application for monitoring Unobtrusive Health |
| **(Henriksson, 2015)** | eHealth Application | Tool for energy balance in children |
| **WPHTP (Yoong, 2015)** | eHealth Application | Application to implement healthy eating and physical training policies |
| **(Singh, 2015)** | eHealth Application | Tool to assist physicians to read radiology reports |

Table 7: Classification of Health Applications based on Disease Control

In following Table, applications related to disease management are gathered.

| Health Application(s) | Health Application Type | Disease Type | Disease |
|---|---|---|---|
| **CHF Tele-management System (Finkelstein J, 2010), EEG-Base Serious Games (Wang, 2010)** | Serious Games | Chronic | Heart related diseases |
| **Match-3 (Scarle, 2011), FatWorld (Sliney A, 2008)** | Serious Games | Chronic | Obesity |
| **Diagnosis and Management of Parkinson (Atkinson, 2010), After Parkinsons Disease (Studios, 2015)** | Serious Games | Chronic | Parkinson Disease |
| **Monster Manor (Boulos, 2015), Empower (Boulos, 2015),Health Seaker (Boulos, 2015)** | Serious Games | Chronic | Diabetes Diseases |
| **Social Skills (Bartolomé, 2010),** | Serious Games | Chronic | Application for neuro-psychological disorder patient |
| **(Sabri H, 2010)** | Serious Games | Chronic | Coronary artery bpass Surgery |
| **(Huckvale, 2015)** | eHealth Application | Chronic | Insuline doze calculation Apps |
| **(Pan, 2015)** | eHealth Application | Chronic | Parkinson Disease |
| **(Ho, 2015), iGetBetter (Zan, 2015)** | eHealth Application | Chronic | Heart Disease |
| **VT for Weight Loss (Arden-Close, 2015)** | eHealth Application | Chronic | Obesity |
| **myWHI (Huguet, 2015)** | eHealth Application | Chronic | Headache in Young Adults |
| **PotM(Lim, 2015)** | eHealth Application | Chronic | Pre-ecllampsia Disorder |
| **Sintromacweb (Ferrando, 2015)** | eHealth Application | Chronic | OAT(Oral Anticoagulation Therapy) |
| **(Volker, 2015)** | eHealth Application | Chronic | Common Mental Disorder |
| **(Cristancho-Lacroix, 2015)** | eHealth Application | Chronic | Alzheimers disease |
| **(Rodriguez, 2015)** | eHealth Application | Chronic | Improve health care in veteran affairs |
| **(Hardinge, 2015)** | eHealth Application | Chronic | Mobile application for monitoring and treatment of COPD(Chr0nic Obstructive Pulmonary Disease) |
| **(Semple, 2015)** | eHealth Application | Acute | Post Operation |
| **ULRFS (Burke, 2009)** | Serious Games | Acute | Upper Libm Rehabilitation |

Table 8: Classification of Health Applications based on Disease Management

# Recommendations

In section 6.1, we discussed recommendations for eHealth challenges related to IT experts. Section 6.2 discussed lessons learned after analyzing different eHealth Applications.

## 6.1 Recommendations for eHealth challenges related to IT Experts

In Information Technology, Ontology is use to describe formal representation to describe concepts and relationship among concepts in a specific domain (Paul Warren, 2006). Ontology is used as a solution to different issues. Researcher used ontology mainly for integrating heterogeneous data, discovering hidden fact, improve the way information is presented, organize and find information for meaning.

Based on these advantages of ontologies, we gave recommendations of following eHealth challenges:

- Detection of disease at early stage.
- Reducing Cost of Health Care by using eHealth System.
- Interoperability of data among different Health Care Places.
- Efficient Managing Patient's Data in eHealth System.
- Effective utilization of Skills of HSR and IT Expert.

Now, we will discuss solution for each eHealth challenge in detail.

### Detection of disease at early stage

Potential solution of this issue is to store and organizes patient data in a format so that we can retrieve information in a way that hidden facts can reveal. Organization of data in form of ontology is possibly best way because as discussed above from ontology we can deduct hidden facts. With already managed knowledge base of evidence and facts of detection of early stage disease can match with electronic patient record data, we can detect disease at its early stage (Matar, 2014).

### Reducing Cost of Health Care by using eHealth System

We can reduce cost of Health care by developing efficient eHealth solutions with help of ontology engineering methodologies. As discussed above, early detection of disease can reduce cost considerably. An effective and efficient eHealth solution which assist Health Practitioner in diagnosing disease correctly may also reduce cost in a way that wrong diagnose can not only increase cost in terms of finance but may also lost of life.

### Interoperability of data among different Health Care Places

Interoperability of data among different Health care places like hospitals, clinics, and basichealth units is a big challenge. As discussed in section 2.7, WHO recommends that data can be interchanged among different Health Practice Places. There are many advantages of sharing data of Patient's record. It may facilitate not only in correct diagnosis but it may also diagnose trend of a specific disease in a particular area/region. As discussed earlier, ontology is used for integrating interoperability of heterogeneous data. Since data of patient record is in heterogeneous form so making interoperability of this heterogeneous patient's record, we can develop ontology using different ontology engineering approaches.

### Efficient Managing Patient's Data in eHealth System

By managing Patient's data efficiently, we can retrieve meaningful information and discover facts. In (Lassere, 2015), authors discussed that how efficient electronic records can useful in improving quality of health care. Ontology is good candidate for data management as we can discover facts efficiently from ontology using different ontology approaches like SPARQL query. In (Moen, 2015;Alnazzawi, 2015), authors suggest semantic technologies can be used to efficiently use electronic patient records.

#### Effective utilization of Skills of HSR and IT Expert

Since time of Health Service Research is very costly, so effective utilization of HSR time with IT expert in helping eHealth solution is a big issue. Instead of creating all rules explicitly or manually by HSR, we can present semi automatic created rules. In that way, his time will reduce by not involving creation of simple rules. Instead, he will utilize his time in creation of complex rules.

## 6.2 Recommendations related to eHealth Applications

From analysis of different Health applications, we identified some socio-economic factors which these applications are providing. We discussed these socio-economic factors in this section and based on these factors, we recommended which type of application should develop so that we imrove quality of Health Care. Next we discussed ecoSystem of Health Application.

### 6.2.1 Socio-economic Benefits of Health Applications

Following socio-economic benefitsare identified during health applications analysis:

- Prevent of loss of Human
- Economic Factor
- Decision Making
- Capacity Building
- Miscellenous Benefits

**Prevent of loss of Human**: PotM (Lim, 2015) is an application in which Pre-ecllampsia Disorder is detected. If this disorder exists in patient and not detected during pregnancy then patient life can be in danger.

**Economic Benefit**: There are some health applications which can reduce cost of Medical care in different ways. Using Parkinson Disease management applications (Pan,2015; Atkinson, 2010; Studios, 2015), patients suffering from this disease needs less visits to patients. Hence, patients saves cost of travelling and Physician conultation fees. It also saves Physician time which is also an economic factor. Physcian can utilise that time in other more productive matters.

**Decision Making**: There are some applications which can help in decision making in different medical processes. These Heart Management Applications (HO,2015;Zan,2015; Finkelstein J, 2010; Wang, 2010)are useful in decision making for physicians.

**Capacity Building**: We analyzed some training and simulation applications which helps physician to increase their technical skills.EMSAVE (HCI Lab, 2011) is a virtul training applcation for emergency medical care. Olive: 3d Hospital Training (Scarle S, 2011), Health Care Monitoring (Fergus, 2009), U-Health Monitoring System (Lee, 2009) and HumanSim (Associates, 2015)are training application related to Hospital Management affairs.Nursing and Midwifery (Skills2Learn, 2015), VI-MED (Mili, 2008) are application for nursing training.

**Miscelleneous Benefits**: Some applications are developed which helps to understand medical education to medical students. Some of such applications are Air Medic Sky (Vtnen A, 2008), Medical Simulation Training Program (Sliney, 2008) and Pulse (BreakWayGames, 2015).WPHTP (Yoong, 2015) is an application to implement healthy eating and physical training policies. Some applicatins are related to body fitness. Some of applications are DDR(Game, 2015), WII (Rego P, 2010) and Re-Mission (Vidani AC, 2010).In (Riazi, 2015), author discussed that eHealth applications are useful for managing diabetes.

Based on above, we recommend that any health application may develop based on above benefits can improve quality of health care.

### 6.2.2 ecoSystem of eHelath Applications

Based on our analysis, we are purposing ecoSystem of an eHealth Application in this section.Efficient patient 's Electronic Medical Record (EMR) management is an integral part of this system. From different larga data management approaches such as ontology development can be use to manage patient's EMR efficiently. From EMR, we can identify which disease is more common in certain area. We can also predict which epidemic disease may approach based on Public Health knowledge. We can priortise which app should develp based on factors discussed in section 6.2.1. After selecting disease, we can sarch an appropriate eHealth app or develop new or update eHealth app. Since Patient and Physician are major stakeholder of this system, so their training can play important role. So, training eHealth applications are developed and also there is need of more training eHealth apps needed for improving quality of Health Care Management.

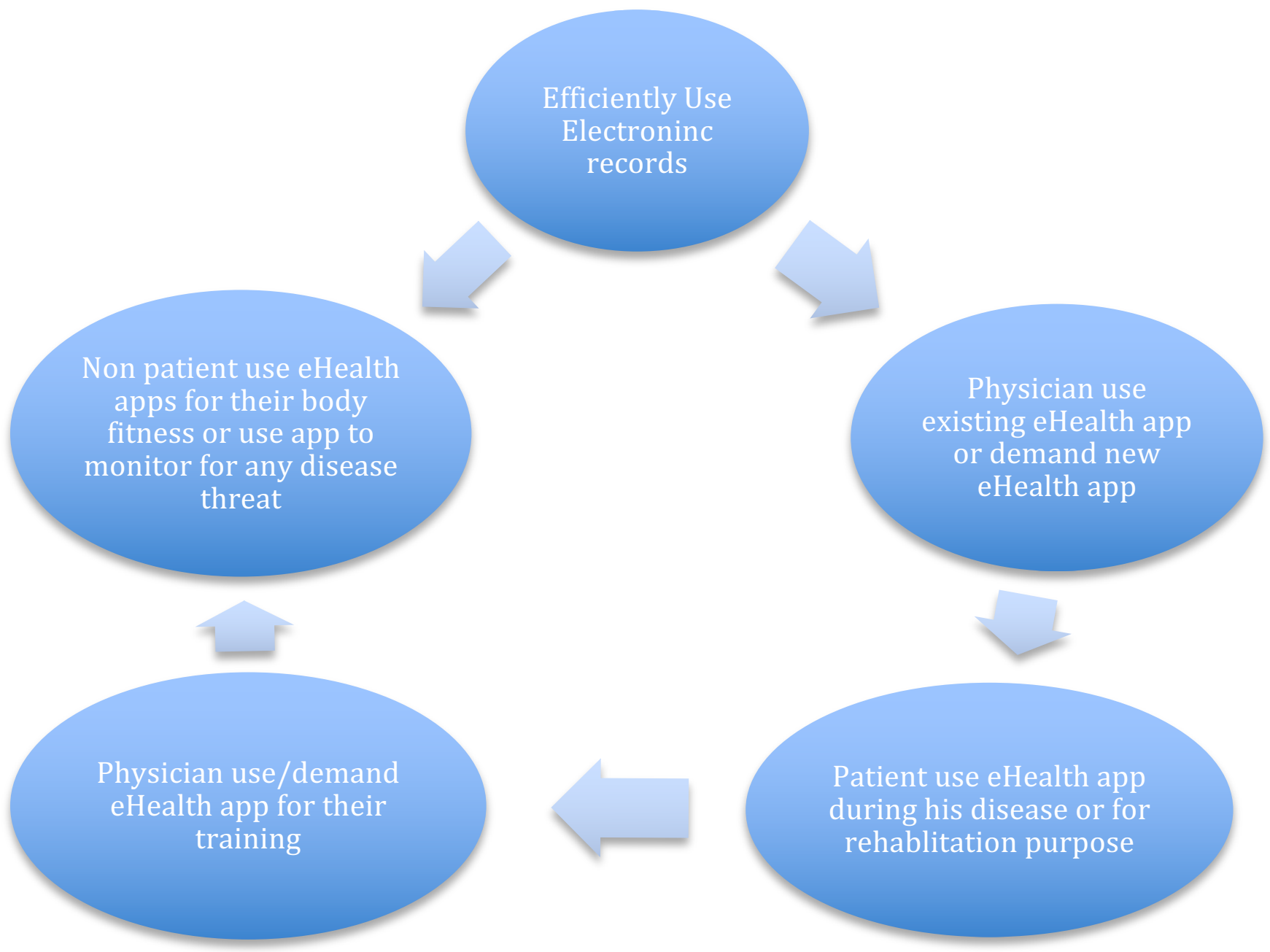

**Figure 6.1: ecosystem of eHealth Application**

# 7 Conclusion and future recommendations

This paper presents comprehensive review of articles related to challenges in eHealth. We categorize these issues in ten different categories. We classify articles based on different parameters in result section. One important result is classifying eHealth issue based on relevant Stakeholder. So, different stakeholders may benefit of this classification. Each stakeholder will concentrate his/her efforts on their relevant problem domain. For example, five issues are related to IT experts. So IT experts can concentrate their efforts on those five issues. Next, we recommend possible solutions to those five key issues briefly. We also presented analysis of recent eHealth applications. We categorize these apps based on different factors. We presented potential socio –economic benefitsif we develop such apps. At the end, we presented ecosystem of eHealth Application.Purpose to show this system is to identify different stakeholders and to identify why and how eHealth application should develop. Thisresearch opens up new avenues for future research in eHealth problems area.We recommend that IT researcher can take future research in each problem area of these five issues discussed in section 5. Researcher in this field can take future research to identify which type of disease can be control or manage from eHealth app and to target which eHealth challenge. For example to target early detection challenge, chronic diseases like cancer, diabetes, and heart diseases are potential candidates. So, development of eHealth apps to control and manage such diseases helps to target early detection eHealth challenge. Similarly researchers can identify which type of disease can be manage and control through serious games.